# An Automatic Mixed Software Hardware Pipeline Builder for CPU-FPGA Platforms

Takaaki Miyajima [†], David Thomas [‡], Hideharu Amano [†]

[†] *Graduate School of Science and Technology, Keio University, Japan*
[‡] *Department of Electrical and Electronic Engineering, Imperial College London, United Kingdom*
vision@am.ics.keio.ac.jp

*Abstract*—Our toolchain for accelerating application called *Courier-FPGA*, is designed for utilize the processing power of CPU-FPGA platforms for software programmers and non-expert users. It automatically gathers runtime information of library functions from a running target binary, and constructs the function call graph including input-output data. Then, it uses corresponding predefined hardware modules if these are ready for FPGA and prepares software functions on CPU by using *Pipeline Generator*. The Pipeline Generator builds a pipeline control program by using Intel Threading Building Block to run both hardware modules and software functions in parallel. Finally, Courier-FPGA dynamically replaces the original functions in the binary and accelerates it by using the built pipeline. Courier-FPGA performs these acceleration processes without user intervention, source code tweaks or re-compilations of the binary. We describe the technical details of this mixed software hardware pipeline on CPU-FPGA platforms in this paper. In our case study, Courier-FPGA was used to accelerate a corner detection using the Harris-Stephens method application binary on the Zynq platform. A series of functions were off-loaded, and speed up 15.36 times was achieved by using the built pipeline.

## I. Introduction

Mixed CPU-FPGA platforms are often used in embedded processing for energy-efficient computing. Critical parts such as computationally intensive functions in a target application are off-loaded to a hardware module which is implemented in reconfigurable logic. On the other hand, non-critical parts run on CPU. By making the best use of the combination of CPU and multiple hardware acceleration modules, a function-level mixed software hardware pipeline should be introduced on the CPU-FPGA platform. For individual hardware modules, methodologies to make implementation easy have been researched, such as high-level synthesis or automatic hardware generation in popular application. Besides, recent application programs often use some open-source libraries such as OpenCV or BLAS. And enough optimized corresponding codes of such open-source libraries are available for popular accelerators like GPUs and FPGAs.

In contrast, automation tools for multiple function/modules management have not been well researched, especially in a mixed CPU-FPGA environment. The demand for such tools is increasing not only by highly trained specialists but also software programmers. We are developing a tool chain for application acceleration, called *Courier-FPGA*, that analyzes a target binary running on the CPU, gathers runtime information of functions and makes a function call graph. It requires only running binaries in this step. And then some functions are off-loaded to predefined corresponding modules in FPGA, and Courier-FPGA builds a function-level pipeline between the software functions and the hardware modules automatically.

In this paper, the technical details of automatic building tool for a mixed software hardware pipeline on CPU-FPGA platforms is presented. The analyzed function call flow are pipelined by our *Pipeline Generator* even if the original binary is not pipelined. We also conducted a practical case study in which Courier-FPGA was used to make a mixed software hardware pipeline on Xilinx's Zynq platform. As a result, an application binary that includes corner detection algorithm were sped up 15.36 times without user intervention.

## II. *Courier-FPGA: A toolchain for Runtime Function Off-load on CPU-FPGA platform*

*Courier-FPGA* [1] is designed for accelerating programs on CPU-FPGA platform for non-expert user. Figure 1 shows an overview and work-step of Courier-FPGA and its three main parts: *Frontend, Courier-FPGA Intermediate Representation (IR), and Backend*. The Frontend first analyzes the running binary and makes function call graph including input-output data, and the Backend automatically off-loaded the functions to the corresponding predefined hardware modules if they ready for FPGA (exist in a database). If a corresponding hardware module is not exist in the database, a function is run on CPU. Hence, the extracted flow is divided into some groups of functions; task. Each task is composed of multiple software functions or hardware modules. Here, a "task" is not a "fine grained" one such as a single x86 assembly code or arithmetic operation on an FPGA, but a process with a certain amount of computation, such as a group of a few functions[2]. In our previous work[1], we described the details of how Frontend analyses the processing flow from the running binary and how Backend replaces the analyzed functions running on the CPU with corresponding functions of a GPU and a FPGA. However current version of Courier-FPGA supports OpenCV and BLAS, it is easy to support another libraries.

### A. Frontend

*Frontend* analyzes a running target binary and makes function call graph including input/output data. This graph includes order of function call, their input/output, and profile log such as processing time, size of input/output data and absolute time of start/end. We used dynamic program analysis and a heuristic approach to make this graph. Users simply start their application as usual, and the Frontend constructs the graph during execution. Although the Frontend doesn't require access to the original source or any sort of re-compilation, it requires an information of a data structure of the target function libraries beforehand. The Frontend is composed of the following three main steps.

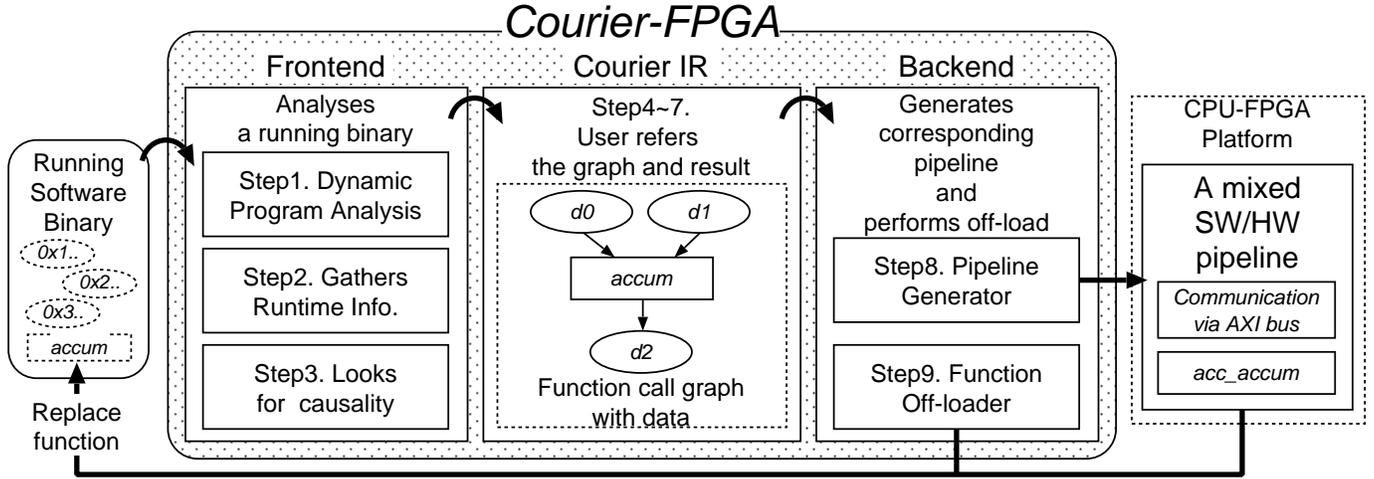

Fig. 1: An overview and work step of the Courier-FPGA: *Frontend* analyses running binary (Step1,2,3) and then constructs a function call graph including input/output data and *Courier-FPGA IR*. User refers the graph and result, decides off-load parts and changes IR if needed (Step 4,5,6,7). And then, *Pipeline Generator* builds a mixed software hardware pipeline (Step 8). Finally, *Function Off-loader* replaces the functions and accelerates the binary on the CPU-FPGA platform (Step 9).

Step 1. The Courier-FPGA traces the running binary by re-ferring the data structure of function libraries,
Step 2. then gathers runtime information during execution,
Step 3. and then looks for the causal function call including input-output data.

### B. Courier-FPGA IR

*Courier-FPGA IR* is an intermediate representation that enables users to modify dataflow and designate parts to off-load to Backend if needed.

Step 4. Courier-FPGA generates an IR corresponding to the processing inside binary
Step 5. then generates the function call graph including input-output data,
Step 6. and then the user examines the graph,
Step 7. after that modifies the processing flow or designates off-load parts if needed.

### C. Backend

*Backend* automatically off-loads the functions, if the corresponding function is ready for a FPGA. *Pipeline Generator* make a mixed software hardware pipeline. and *Function Off-loader* in Backend automatically optimizes data communication along with off-load, and maintains original processing flow before and after off-load. We explain the technical details of Pipeline Generator in the next section.

Step 8. The Pipeline Generator generates the corresponding hardware module on an FPGA, and then builds a mixed software hardware pipeline.
Step 9. Finally, the Function Off-loader selects a path and replaces functions with the generated pipeline

An example of processing step of Courier-FPGA is illustrated in Figure 1. Here, running software binary contains a processing flow, "accum" function, which obtains two input data (d0 and d1) and produces one output data (d2). Then "accum" is replaced with corresponding accelerator function "acc_accum" and off-loaded. A caption of the figure describes the work-flow of Courier-FPGA. The user can refer a function call graph including input/output data and modify it in Courier-FPGA IR interactively.

## III. AN AUTOMATIC MIXED SOFTWARE HARDWARE PIPELINE BUILDER

In Step 8, the *Pipeline Generator* automatically searches corresponding predefined hardware modules from a database by functions name, places them on FPGA and prepares software functions. Then, it makes a software program that runs a mixed software hardware tasks in parallel so as to make the best use of the parallelism. The parallel tasks perform processing which corresponds to a target binary. In Step 9, *Function Off-loader* wraps the built pipeline and actually replaces with the function in the target binary and the pipeline is ready to deploy. During deployed run, the control program made by the Pipeline Generator starts to accelerate the binary. Even if the functions in the target binary run sequentially, Courier-FPGA can run them in a pipelined manner.

### A. Structure of a mixed software hardware pipeline

Structure of a mixed software hardware pipeline on a CPU-FPGA platform is composed of the following three main parts,

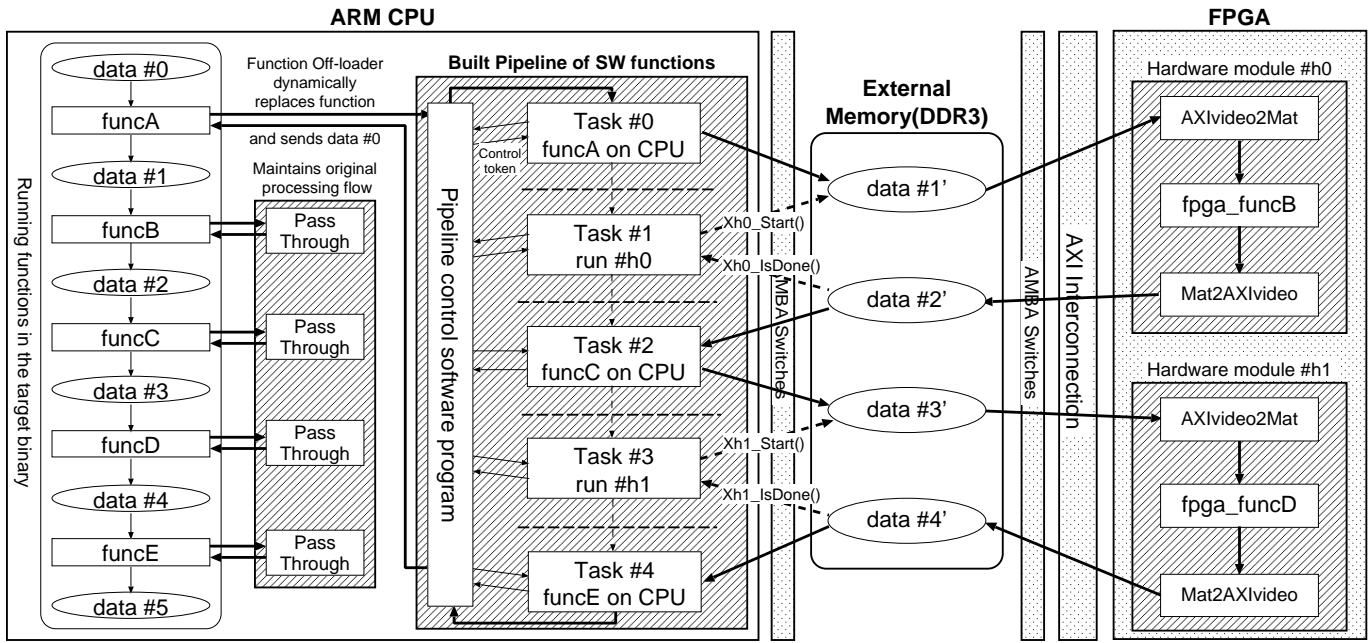

Fig. 2: Behavior of a mixed software hardware pipeline controlled by software program during deployed run. Shaded rectangles are generated by Pipeline Generator and bold lines explain data flow. Tasks run in a pipelined manner, and each task can send and receive input and output data which is indicated by bold line. Input and output data of tasks are stored in the external memory. In this case, Task #1 and #3 run hardware module #h0 and #h1 while Task #0, #2 and #4 run software function.

- A pipeline control software program that controls the software hardware tasks.
- Software functions run on the CPU.
- Hardware modules run on the FPGA.

The control program is needed in order to run multiple software and hardware functions in a pipelined manner. Software functions are used when corresponding hardware modules doesn't exist in a database.

In Figure 2 the Pipeline Generator generates five-stage pipeline, two hardware modules and three software functions. The pipeline control program runs these tasks in a pipelined manner. On a deployed run, tasks work as follows from the viewpoint of the target binary. The Function Off-loader hooks funcA and its input data (data #0) from the running functions in the target binary, which is illustrated on the left of the figure. Then, Task #0 processes funcA and stores the result (data #1') in an external memory. And then, Task #1 invokes "start command" (Xh0_Start()) to send the data to the hardware module #h0, and receives "done signal" (Xh0_Done()) when fpga_funcB finishes and stores a result data (data #2') in the memory. The second input data from the running binary are simultaneously processed by Task#0. This is a software controlled task pipeline. Note that intermediate data such as "data #1'" are stored in the external memory.

*B. Building a mixed software hardware pipeline*

Figure 3 shows a processing flow of how the Courier-FPGA automatically builds a pipeline. First of all, the Frontend analyzes a target binary and makes a list of running functions name. Then, the Backend searches corresponding modules from a hardware module database. Found corresponding modules are synthesized and placed on FPGA. On the other hand, functions which cannot be found becomes software ones run on CPU. Finally, Pipeline Generator builds a balanced pipeline considering processing time of functions/modules. Each stage of the pipeline runs hardware module(s) on FPGA or software function(s) on CPU.

In the case of Figure 2, the Backend found two corresponding hardware functions: funcB and E in the database. Then, Pipeline Generator generates two hardware modules: the former contains fpga_funcB, and the latter contains fpga_funcD. In addition, Task #1 and Task #3, a software part of these modules, were also prepared. Software parts performs communication between hardware modules and an external memory. On the other hand, there were no hardware modules for funcA, C and E in the database, so software functions are used for them. Thus, five tasks, two hardware modules and three software functions, are used for the five-stage pipeline. The pipeline control program runs these tasks in a pipelined manner and keeps the order of stages.

*1) Generating hardware modules:* As we described above, the Backend searches a corresponding hardware module in a database for each analyzed function. Technically, this database includes an OpenCV-compatible high-level synthesis library provided by Xilinx[3]. The Pipeline Generator generates codes of the hardware modules, input/output port and optimization pragma. For the hardware modules, Pipeline Generator simply generates the corresponding module name. For example, *hls::Sobel* is used for *cv::Sobel* as a corresponding function and arguments are defined. For the input/output port, Pipeline

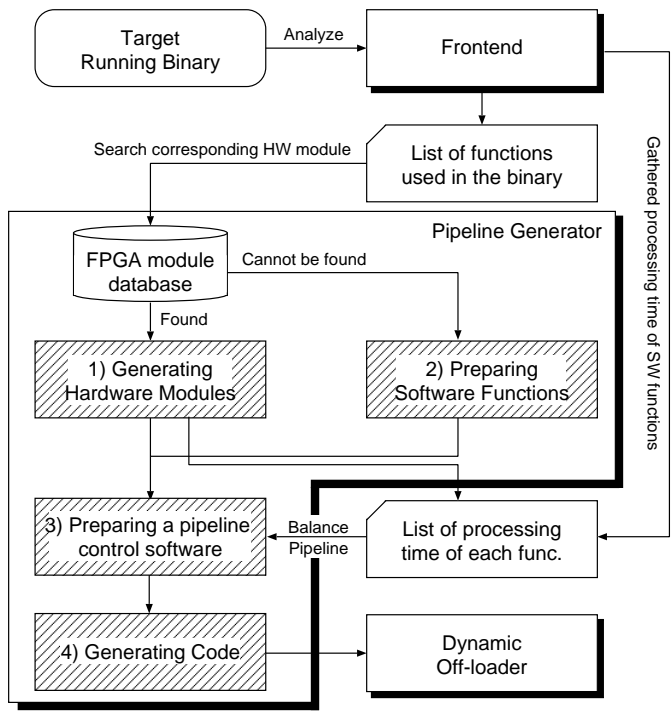

Fig. 3: A processing flow of building a mixed SW/HW pipeline. Shadowed rectangles are parts of the Courier-FPGA and ones filled with oblique lines are explained in Sect.III B.

Generator uses *AXIvideo2Mat* and *Mat2AXIvideo*. These are the code of AXI4-Streaming protocol including Video DMA controller that communicates the CPU and the hardware module. For the optimization pragma, Pipeline Generator inserts *#pragma HLS dataflow* in default in order to achieve shorter processing time.

In the case of Xilinx's OpenCV library, each function is optimized for per pixel processing. In addition, input data from the software is first stored in the DDR3 on-board RAM on Zynq before being processed and stored again in the RAM after processing. This kind of streaming architecture requires to read and write the data into the DDR3. Hence, the bus width of the input and output port significantly influences the performance. The Pipeline Generator automatically calculates and defines the width by using the extracted bit-depth information from the Frontend. Furthermore, the Pipeline Generator tries to pipeline a series of functions if the functions have no branch nor loop. This pipelining is performed by inserting *#pragma HLS STREAM* in the head of the generated functions.

Generated hardware modules are prepared as a block device on Linux, and a basic device driver APIs that send/receive data are prepared by Xilinx's high-level synthesis tool. In the case study, *XTask0_Start()* function sends input data to start process on hardware module, and *XTask0_IsDone()* function polls done signal until the hardware module finishes a process. These API functions are used in a task on CPU side.

*2) Preparing Software functions:* For each software function task, the Backend prepares to run original functions in the binary and dynamically replaces during deployed run. The Backend first looks for a function name in function libraries by using *dlsym* [4] with "RTLD_NEXT" option so as to use an original function. The functions library is designated by *dlopen* with "RTLD_LAZY" option [4]. The reason why we used such method is that Dynamic Off-loader basically uses a software technique called *DLL injection* so as to realize dynamic off-loading. When we use DLL injection and want to use software functions in the target binary, the above described method is required. In the case of hardware modules, we don't need to use this method.

*3) Preparing a pipeline control software:* For a pipeline control software program, we use Intel Thread Building Blocks (TBB) that runs multiple functions in a pipelined manner on CPU. The *tbb::pipeline* class is provided to build a pipeline. A user can add an arbitrary task to each stage of the pipeline skeleton, and specify the processing order of the stages. After that, TBB automatically runs the tasks in a pipelined manner. TBB introduces the concepts of a thread pool and token base pipeline. Multiple slave threads are managed by a master thread and user write codes for each slave thread (e.g. software functions or hardware modules). TBB is also capable of double buffering when two or more tasks are running.

Unlike a common hardware pipeline in which the previous stage cannot start until the next stage has finished, a pipeline provided by TBB can start each stage even if the next stage doesn't finish. For example, Task #0 can take the second input while Task #1 is processing a time consuming task for the first input. As a result, the pipeline can reduce the probability of stall compared with the hardware pipeline. Additionally, stages which run in parallel can be dynamically changed since an idle thread is randomly chosen by the control program.

*4) Generating Code:* We use a Python and Jinja2 [5] to implement the Pipeline Generator. The Pipeline Generator is a script and technically composed of a pipeline skeleton part and a task part for each stage. The former is almost static and the latter is flexible since it contains above described software functions and hardware modules. The Backend tells information (e.g. a filter type of Intel TBB, data type of input/output or actual processing code) to the Pipeline Generator, and it generates the whole code.

Current Pipeline Generator divides the extracted processing flow into some stages by using the simple partitioning policy: "Pipeline Generator divides total processing time by the number of thread plus one and searches the closest sub-total of processing time of functions". The policy is derived from the following considerations. According to our preliminary evaluation, the number of stages should close to that of a logical thread of the Zynq (= 2). This is because the control load of master thread and the communication frequency of intermediate data should be reduced for a streaming architecture. Furthermore, to keep the minimal processing time, each pipeline stage should run in nearly the same time, i.e. a balanced pipeline. Note that, processing time of software functions can be obtained in the analyzed data from the Frontend and that of hardware modules can be estimated by the logic synthesis tool, and thus processing time of all functions are available. Additionally, the Pipeline Generator defines the first and last functions define serially run (*serial_in_order*), while the rest of the function run in parallel (*parallel*).

## C. Off-loading tasks

The *Function Off-loader* in the Backend automatically makes a wrapper to replace the original function in the binary. The wrapper contains a generated code by Pipeline Generator, *Off-loader Switcher* and some pre/post-processing and data transfer. Finally, generated code is compiled as shared object and it is replaced to the running binary during deployed run. This technique is known as DLL injection.

## IV. CASE STUDY

We show a details process of Courier-FPGA by describing a case study. The experimental conditions were as follows: the running binary was analyzed on MacOS X 10.9 64bit, The binary was deployed on Zynq-7000 AP SoC (XC7Z020-CLG484-1) on Zedboard. Zynq-7000 was composed of a Dual Core ARM Coretex-A9 CPU 667MHz with 512MB memory and 85,000 Artix-7 programmable logic cells. Linaro 32bit (Debian 7.0) ran on the ARM CPU. We also used Xilinx Vivado HLS and Vivado 2013.2 as a synthesis tool.

### A. cornerHarris; a corner detection application

"cornerHarris_Demo" is a sample program of corner detection that is contained in OpenCV (opencv-2.4.x/samples/cpp/tutorial_code/TrackingMotion/cornerHarris_Demo.cpp). It demonstrates corner detection using the Harris-Stephens method [6] written by using OpenCV; a widely used open software library for computer vision. A binary of cornerHarris_Demo was mainly composed of four functions listed in Table I. And we inputted a 1920 x 1080 size image. Note that we ignored *imread* that reads the first data since it is executed only once.

Figure 4 is a graph of function call including input/output data constructed by Frontend. Ellipse nodes and rectangle nodes represent images and functions, respectively. The size of the node reflects the execution time or the size of the data. The processing time is shown in the second row of the rectangle node and the size of the data is shown in that of the ellipse node. (height × width × bit-depth × channels; e.g., the first node is 1920 × 1080 × 24bit × 1-channel). Nodes are aligned in chronological order so that functions run in a sequential manner. According to the graph, it processed in 1371.1 [ms] in total. In addition, *cv::cornerHarris* and *cv::convertScaleAbs* comprise 65% and 15% of the whole processing time respectively.

Courier-FPGA built a four-stage pipeline and shortened the processing time to 83.8[ms] or x15.36 speedup compared with the original binary. In Table I, "Original Binary" indicates the target binary running on the ARM CPU, and "Courier-FPGA" indicates a result of acceleration that uses both CPU and FPGA. AXIvideo2Mat is input to the hardware module via AXI bus and Mat2AXIvideo is vice versa.

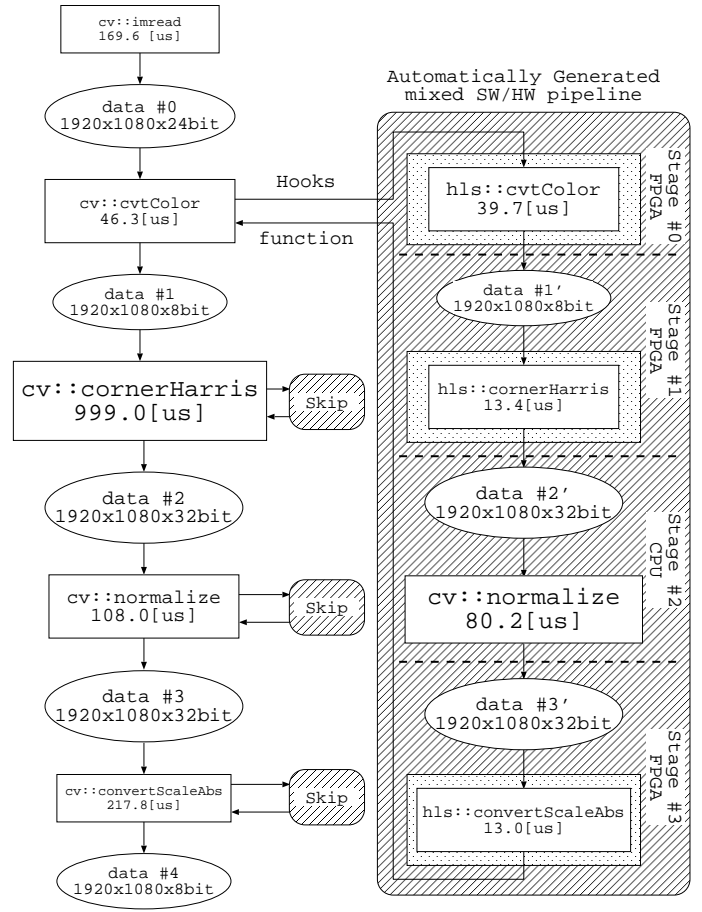

Fig. 4: Processing flow analyzed from the running binary (left) and off-loaded flow (right). Each processing step is assigned to task. The Pipeline Generator generates a four-stage mixed software hardware pipeline, and Function Off-loader hooks the head of four functions during deployed run.

Tables II and III show evaluations of the modules generated for Stage #0, #1 and #3. The hardware speeded up by 1.16 times, by 73.5 times and by 15.8 times resplectively. These time include data communications via the AXI Stream bus. In the hardware module generation step, Pipeline Generator first tried to make *cvtColor* and *cornerHarris* into single hardware module. Although generated module was too slow to use.

TABLE I: Processing time comparison ([ms])

|  | Original Binary | Courier-FPGA | Running on |
|---|---|---|---|
| cvtColor | 46.3 | 39.8 | FPGA |
| cornerHarris | 999.0 | 13.6 | FPGA |
| normalize | 108.0 | 80.2 | CPU |
| convertScaleAbs | 217.8 | 13.2 | FPGA |
| Total | 1371.1 | 83.8 | CPU&FPGA |
| Speed-up | x1.00 | x15.36 | — |

## V. RELATED WORK

There has been an enormous amount of research on hardware-software co-design for reconfigurable systems. Most of them refer source code and sophisticated compilers or HLS techniques unlike Courier-FPGA, which performs pipelining on a CPU-FPGA platform by using predefined HW modules.

TABLE II: Evaluation: Synthesis of individual module

| Module | Freq. [MHz] | Latency [clk] | Proc. time [ms] |
|---|---|---|---|
| hls::cvtColor | 157.2 | 6,238,090 | 39.7 |
| hls::cornerHarris | 157.9 | 2,111,579 | 13.4 |
| hls::convertScaleAbs | 160.6 | 2,090,882 | 13.0 |

TABLE III: Evaluation: Resource utilization of modules

| Module | BRAM | DSP48E | FF | LUT |
|---|---|---|---|---|
| Stage#0: hls::cvtColor | | | | |
| Sub total | 23(8%) | 10(4%) | 4013(3%) | 5550(10%) |
| AXIvideo2Mat | 0 | 0 | 195 | 237 |
| hls::cvtColor | 23 | 10 | 3631 | 4343 |
| Others | 0 | 0 | 187 | 970 |
| Stage#1: hls::cornerHarris | | | | |
| Sub total | 66(23%) | 15(6%) | 13596(12%) | 17494(32%) |
| AXIvideo2Mat | 0 | 0 | 92 | 133 |
| hls::cornerHarris | 66 | 15 | 12869 | 14881 |
| Mat2AXIvideo | 0 | 0 | 58 | 109 |
| Others | 0 | 0 | 577 | 2371 |
| Stage#3: hls::convertScaleAbs | | | | |
| Sub total | 0(0%) | 0(0%) | 1195(1%) | 2307(4%) |
| AXIvideo2Mat | 0 | 0 | 92 | 133 |
| hls::convertScaleAbs | 0 | 0 | 920 | 1805 |
| Mat2AXIvideo | 0 | 0 | 58 | 109 |
| Others | 0 | 0 | 125 | 260 |
| Total | 89(31%) | 25(10%) | 18804(16%) | 25351(46%) |

Also, there are a number of researches targeting binary translation [7] [8] [9] [10] [11]. Most of these researches focus on analyzing and translating instruction-level behavior into hardware circuits. Stitt et al. proposed *Warp Processing*, which takes advantage of the reconfigurability of the FPGA [7] [8]. Their unique points are the original CAD module, which analyzes the code to detect hot-spots, and automatic generation of FPGA circuits. Bispo et al. proposed hardware-based instruction bus profiling to measure the branch frequency of loops (*Megablock*)[9]. Although Megablock can offload a large block with a number of instructions, it requires special hardware for profiling and off-loading. Nathan et al. presented the *Configurable Compute Array* (CCA) for automatically designing new instruction sets by using an FPGA as a substitute for a series of existing operations on the CPU[10]. They also show how to determine the operations by using both a dynamic profile and static one. Other researches on automatic transformation of assembly language to hardware modules has been done. For example, *eMIPS*[12], *Binary-translation Optimized Architecture(BOA)*[13] and *Dynamic Instruction Merging* (DIM)[11]. These studies try to convert the basic blocks in a software binary into a hardware module, and proposed specific means for doing so.

Unlike the above studies, the target of Courier-FPGA is a coarse-grained dataflow and focuses on building a mixed software hardware pipeline with the corresponding HDL description of the target functions. It also cosidering a data transfer between modules. Courier-FPGA assumes that the corresponding HDL description of the target function is ready. Thus, pipelining of Courier-FPGA can be combined with traditional HLS techniques or binary translation techniques which focus on acceleration of individual functions.

## VI. CONCLUSION

This paper presented the technical details of *Courier-FPGA*'s *Pipeline Generator*; an automatic mixed software hardware pipeline builder for CPU-FPGA platforms. It prepares software functions, uses predefined hardware modules and makes a pipeline control software program in order to run software and hardware tasks in parallel. Courier-FPGA dynamically replaces the original functions in the binary and accelerates it by using the built pipeline. We demonstrate our toolchain by describing a practical case study we conducted on a corner detection algorithm on the Zynq platform. 15.36 times speedup was achieved without user intervention.

In our future work, we will investigate how Courier-FPGA handles more complicated processing flow which includes data dependency or control flow.

## VII. ACKNOWLEDGMENT

The present study is supported in part by the JST/CREST program entitled "Research and Development on Unified Environment of Accelerated Computing and Interconnection for Post-Petascale Era" in the research area of "Development of System Software Technologies for post-Peta Scale High Performance Computing".